\documentclass[aps,prl,superscriptaddress,preprint]{revtex4}
\usepackage{amsmath,epsfig}
\usepackage{graphicx}

\def\be{\begin{equation}}
\def\ee{\end{equation}}
\def\bea{\begin{eqnarray}}
\def\eea{\end{eqnarray}}
\newcommand{\ket}[1]{\mbox{$|#1\rangle$}}
\newcommand{\bra}[1]{\mbox{$\langle#1|$}}

\def\GammaD{\Gamma_{\footnotesize\textrm{1D}}}

\begin{document}

\title{Cavity QED with atomic mirrors}
\date{\today}

\author{D.E. Chang}
\affiliation{ICFO - Institut de Ciencies Fotoniques, Mediterranean
Technology Park, 08860 Castelldefels (Barcelona), Spain}

\author{L. Jiang}
\affiliation{IQIM, California Institute of Technology, Pasadena,
CA 91125, USA}

\author{A.V. Gorshkov}
\affiliation{IQIM, California Institute of Technology, Pasadena,
CA 91125, USA}

\author{H.J. Kimble}
\affiliation{IQIM, California Institute of Technology, Pasadena,
CA 91125, USA} \affiliation{Norman Bridge Laboratory of Physics
12-33, California Institute of Technology, Pasadena, CA 91125,
USA}

\begin{abstract}
A promising approach to merge atomic systems with scalable
photonics has emerged recently, which consists of trapping cold
atoms near tapered nanofibers. Here, we describe a novel technique
to achieve strong, coherent coupling between a single atom and
photon in such a system. Our approach makes use of collective
enhancement effects, which allow a lattice of atoms to form a
high-finesse cavity within the fiber. We show that a specially
designated ``impurity'' atom within the cavity can experience
strongly enhanced interactions with single photons in the fiber.
Under realistic conditions, a ``strong coupling'' regime can be
reached, wherein it becomes feasible to observe vacuum Rabi
oscillations between the excited impurity atom and a single cavity
quantum. This technique can form the basis for a scalable quantum
information network using atom-nanofiber systems.
\end{abstract}
\maketitle

Techniques to controllably interface atoms with quantum optical
fields form the basis for many applications in quantum information
science~\cite{duan08,kimble08}. For example, photons are
convenient to relay information over large quantum networks, while
atoms naturally are physical systems that can process and store
this information. Thus far, the available techniques to
efficiently couple single photons with atomic media fall into one
of the following, mostly independent, categories: i) cavity
quantum electrodynamics~(QED)~\cite{miller05,walther06,haroche06},
where atomic interactions with light are enhanced via a
high-finesse cavity, ii) coherent coupling with atomic ensembles
exhibiting large optical depths~\cite{hammerer10}, and iii) the
use of fields tightly focused to dimensions smaller than or
approaching the scattering cross-section of a single
atom~\cite{vanenk01,darquie05,wrigge08,chang06,akimov07,tey08,hetet11}. Although
remarkable achievements have been made with all of these
approaches, a robust, scalable technique that can be easily
integrated with photonics remains elusive.

Here, we describe a hybrid strategy that combines appealing
attributes of each of the methods described above, and which can be implemented with relatively modest resources. Our
approach utilizes a promising atom-light interface developed in
recent years, which consists of cold atoms trapped near tapered
nanofibers~\cite{nayak07,vetsch10}. The traps are
well-characterized~\cite{nayak07,vetsch10,lacroute11} and can
potentially be used to transport and couple atoms to other
systems, such as dielectric optical
cavities~\cite{aoki06,colombe07,alton11} and nanomechanical
resonators~\cite{treutlein07,hammerer09}. The nearly
diffraction-limited transverse confinement of optical fields thus
far enables $\sim 10\%$ coupling efficiency of a single atom to
the fiber~\cite{nayak07,vetsch10}, which has allowed for
observations of strong light-matter interactions using relatively
few atoms and low powers~\cite{renn95,londero09,bajcsy09}.

Our hybrid approach is based upon the following principles. First,
we show that although the single-atom coupling in this system
might be relatively weak, there exist collective modes of a
trapped atomic ensemble whose coupling to light is enhanced by the square
root of the atom number, $\sqrt{N_A}$~\cite{hammerer10}. While
collective effects are generally well-known, special consequences
emerge in the nanofiber system when the atoms are trapped in a
lattice. In particular, collective effects cause such a lattice to
act as a near-perfect mirror for an incident field close to
resonance. In analogy to cavity QED, we then demonstrate that two
sets of atomic mirrors can form an effective cavity, which can
greatly enhance the coupling of a \textit{single}, specially
chosen ``impurity'' atom~(or a few impurity atoms) positioned
inside. We introduce a novel quantum spin model to describe the
atom-light coupling, which allows one to exactly map the
atom-nanofiber interface onto the simple and elegant
Jaynes-Cummings model of cavity QED~\cite{jaynes63}. A unique feature of our atomic mirrors compared to conventional cavities is that they have long relaxation times and are highly dispersive. Remarkably, even with very low mirror finesse~($F\sim 10^2$) this property allows one to attain the ``strong coupling'' regime of cavity
QED~\cite{miller05,walther06,haroche06}, where vacuum Rabi
oscillations~\cite{sanchez-mondragon83,thompson92,brune96,boca04}
occur between an excited impurity atom and a single ``photon''
stored in the cavity~(or more precisely, in the atomic mirrors). Furthermore, as quantum mechanical objects, these atom mirrors can be used to store quantum information and transfer this information into propagating waveguide modes. We describe how these various features can be combined to realize all of the building blocks for scalable quantum information processing.

\section{Results}
\subsection{Atom-nanofiber interface: linear spectral properties}

\begin{figure}[p]
\begin{center}
\includegraphics[width=11cm]{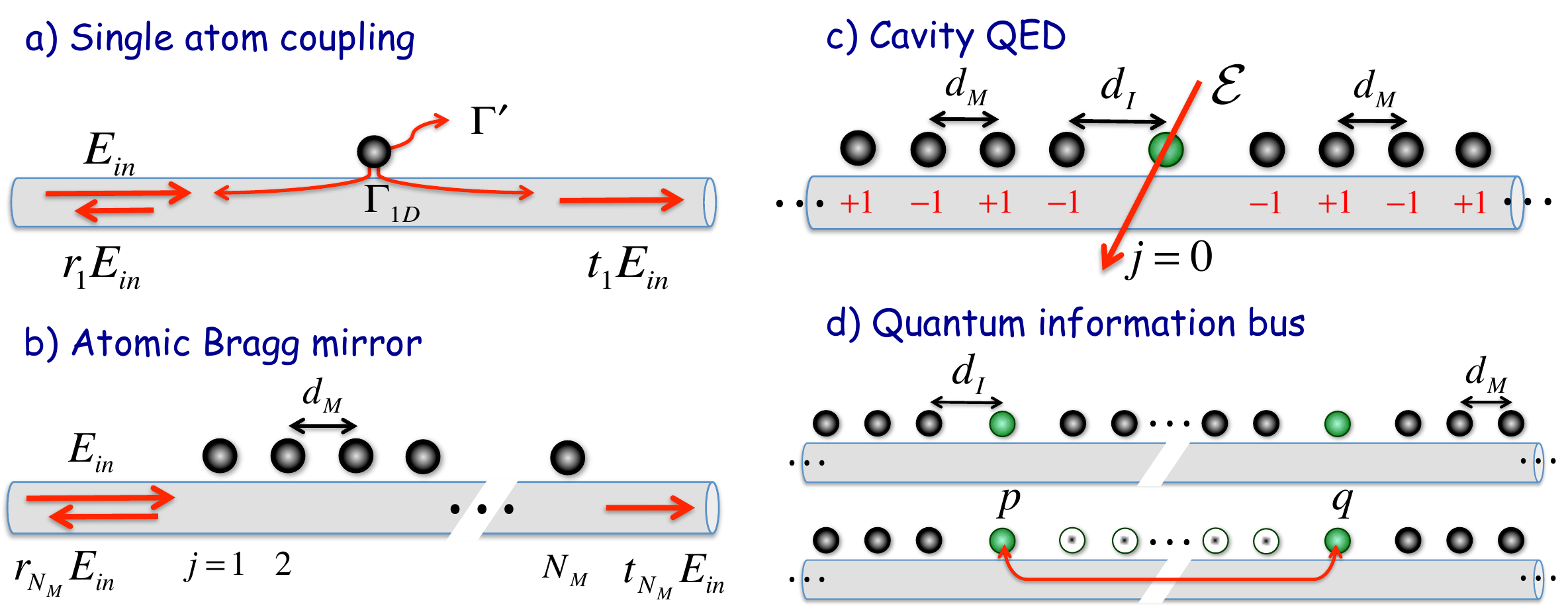}
\end{center}
\caption{\textbf{Different configurations of a coupled
atom-fiber system.} \textbf{a)} Single atom coupling. The atom spontaneously
emits into the fiber and free space at rates $\GammaD$ and
$\Gamma'$, respectively. In the linear regime, the atom scatters a
guided input field $E_{\footnotesize\textrm{in}}$ with reflection
and transmission amplitudes $r_1,t_1$. \textbf{b)} $N_M$ atoms in a chain
with lattice constant $d_M$ form an atomic ``Bragg mirror,'' with
linear reflection and transmission amplitudes $r_{N_M},t_{N_M}$.
\textbf{c)} In the ``cavity QED'' configuration, two atomic Bragg
mirrors~(located at $1\leq j \leq N_M$ and $-N_M\leq j \leq -1$)
form a cavity, which enhances the coupling of an impurity
atom~(green, $j=0$) to the fiber. The distance between the
impurity and its nearest neighbors is $d_I$. The relative phases
$\pm 1$ of the mirror atom spin wave comprising the cavity
excitation are denoted in red. An external field $\mathcal{E}$ can
be used to drive the impurity atom. \textbf{d)} Quantum information
transfer can occur between two well-separated impurity atoms $p,q$
in the ``quantum information bus'' configuration. Here the two
impurity atoms initially sit in separate cavities within a long
chain of mirror atoms~(dark circles). Then, all the mirror atoms
between them are flipped into a transparent hyperfine state
$\ket{s}$~(white). This process loads the impurity atoms into a
new, common cavity mode defined by the remaining mirror atoms
positioned external to $p,q$.\label{fig:chain}}
\end{figure}

We model our system as an ensemble of two-level atoms with ground
and excited states $\ket{g},\ket{e}$ and resonance frequency
$\omega_A$, located at positions $z_j$. These atoms interact with
a one-dimensional waveguide supporting left- and right-propagating
fields $\hat{E}_{L,R}$ with linear dispersion and velocity $v$
through a dipolar coupling,
$H_{\footnotesize\textrm{int}}=-\hbar{\beta}\sqrt{2\pi}\sum_{j}\left[\sigma_{eg}^{j}(\hat{E}_{R}(z_j)+\hat{E}_L(z_j))+h.c.\right].$
This coupling yields the Maxwell-Bloch equations for the field
propagation~\cite{chang07b} with solutions
\be
\hat{E}_{R(L)}(z,t)=\hat{E}_{R(L),\footnotesize\textrm{in}}(z\mp
vt)+\frac{\sqrt{2\pi}i\beta}{v}\sum_{j}\Theta\left(\pm(z-z_j)\right)\sigma_{ge}^{j}(t\mp(z-z_j)/v),\label{eq:fieldsoln}
\ee
where $\Theta(z)$ is the Heaviside step function. The single-atom
spontaneous emission rate into the waveguide is $\GammaD=4\pi
\beta^2/v$~\cite{chang07b}. In addition to
equation~(\ref{eq:fieldsoln}), $H_{\footnotesize\textrm{int}}$ yields
the usual Heisenberg equations for the atomic coherence operators
$\sigma_{ge}^j=\ket{g_j}\bra{e_j}$. We also assume that each atom
independently emits into free space with rate $\Gamma'$, such that
the total emission rate of a single atom is
$\Gamma=\Gamma'+\GammaD$~(see figure~\ref{fig:chain}a).

The transfer matrix formalism of Ref.~\cite{deutsch95} is
well-suited to solve for linear or single-photon propagation
through the ensemble. From equation~(\ref{eq:fieldsoln}), one first
finds the single-atom reflection and transmission amplitudes of an
incident field~\cite{chang07b}, as shown in figure~\ref{fig:chain}a.
We find that $r_1(\Delta_A)=-\GammaD/(\Gamma-2i\Delta_A)$ and
$t_1(\Delta_A)=1+r_1(\Delta_A)$, where
$\Delta_A=\omega_P-\omega_A$ is the detuning between the field
input frequency $\omega_P$ and the atomic resonance. In addition,
free-space propagation over a distance $d$ is characterized by
multiplicative phase shifts,
$E_{R(L)}(z+d)=e^{{\pm}i\omega_{P}d/v}E_{R(L)}(z)$. The field
scattering from many atoms can then be exactly summed using
transfer matrices~\cite{deutsch95}, from which the total system
reflection and transmission amplitudes are obtained.

\begin{figure}[p]
\begin{center}
\includegraphics[width=11cm]{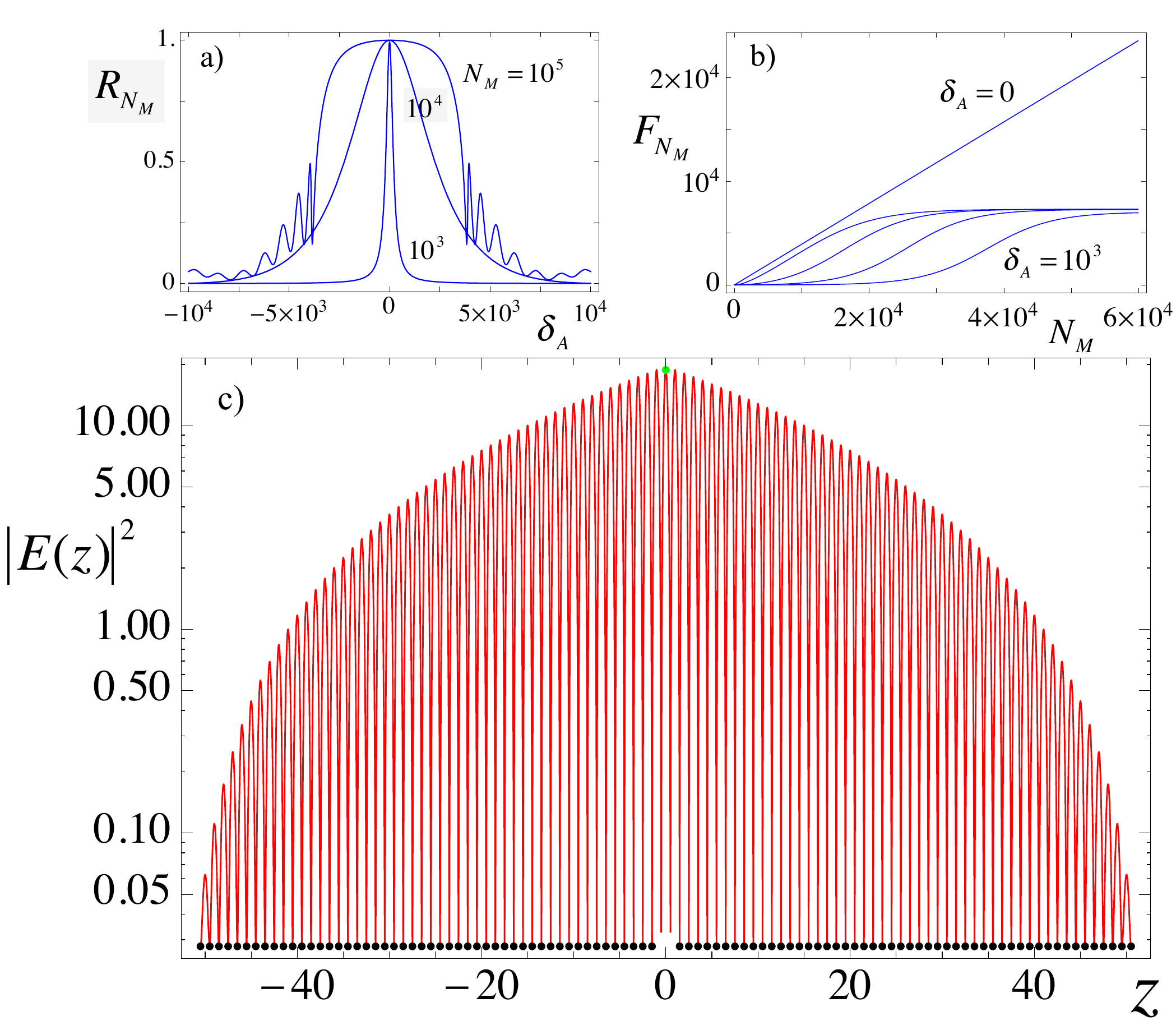}
\end{center}
\caption{\textbf{Atom mirror properties.} \textbf{a)} Reflectance
$R_{N_M}=|r_{N_M}(\delta_A)|^2$ of a mirror comprised of an atomic chain, as a function of dimensionless
detuning~$\delta_A=\Delta_A/(\Gamma/2)$. The spectra are shown for
mirror atom numbers $N_M=10^3,10^4,10^5$. The reflectance becomes
non-Lorentzian for atom numbers $N_M\gtrsim
N_{\scriptsize\textrm{gap}}\equiv\sqrt{\omega_A/\GammaD}\approx
1.6 \times 10^4$. \textbf{b)} Effective cavity finesse, defined
as $F_{N_M}\equiv\pi/(1-R_{N_M}(\delta_A))$, of an atomic chain as
a function of mirror atom number $N_M$. The finesse is shown for
detunings $\delta_A=0,30,100,300,1000$. \textbf{c)} Two atom mirrors surrounding an impurity atom form an effective cavity, as illustrated in figure~\ref{fig:chain}c. The intra-cavity intensity $|E(z)|^2=|E_{R}(z)+E_{L}(z)|^2$ is plotted
as a function of position~(in units of the atomic site number
$j$), when the impurity atom is externally driven on resonance. $|E(z)|^2$ is normalized by the
intensity produced by a single atom driven on resonance under the
same external amplitude $\mathcal{E}$, in the absence of mirror
atoms. The black and green points depict the local fields at the
mirror and impurity atom sites, respectively. We have used
parameters $\GammaD=\Gamma'/4$ and $\omega_A/\Gamma=5.4\times
10^7$ for all panels in this figure.\label{fig:cavity}}
\end{figure}

We now focus on the case where $N_M$ atoms are arranged
periodically with a lattice constant of $d_M=\pi v/\omega_A\equiv
\lambda_A/2$ to form an atomic ``Bragg mirror,'' as shown in
figure~\ref{fig:chain}b~(analogous results hold when $d_M$ is any
other integer multiple of half the resonant wavelength
$\lambda_A$). For atom number $N_M\lesssim
N_{\footnotesize\textrm{gap}}\equiv\sqrt{\omega_A/\GammaD}$, the
effect of small detunings from resonance is negligible in free
propagation, and one can approximate
$e^{{\pm}i\omega_{P}d_M/v}\approx -1$. The reflectance from the
lattice in this regime is given by a broadened Lorentzian,
$R_{N_M}(\Delta_A)=\frac{(N_M\GammaD)^2}{(\Gamma'+N_M\GammaD)^2+4\Delta_A^2}$~(see
figure~\ref{fig:cavity}a), while the transmittance is
$T_{N_M}(\Delta_A)=\frac{\Gamma'^2+4\Delta_A^2}{(\Gamma'+N_M\GammaD)^2+4\Delta_A^2}$.
For $N_M\gtrsim N_{\footnotesize\textrm{gap}}$, the resonant
reflectance approaches unity with increasing atom number,
$R_{N_M}(\Delta_A=0)=\left(\frac{N_M\GammaD}{\Gamma'+N_M\GammaD}\right)^2$,
but the phases accumulated in free propagation for finite detuning
cannot be neglected. Away from resonance, the lattice forms a band
gap for detunings $|\Delta_A|<\sqrt{\omega_A\GammaD/\pi}$, over
which the reflectance saturates as $N_M\rightarrow\infty$ to a
value
$1-R_{N_M}\sim\mathcal{O}(\sqrt{\Gamma'^2/(\omega_A\GammaD)})$.
The deviation from perfect reflection occurs because of atomic
scattering of light into free space, in contrast to the perfect
gap formed by purely dispersive media. Similar results have been
derived for the present geometry~\cite{changy11} and for atoms
trapped in a one-dimensional standing wave in free
space~\cite{deutsch95}, as well as observed in the latter
case~\cite{birkl95,schilke11}. Band structures in other atomic
configurations have also been explored~\cite{petrosyan07,zoubi10}.
In the following, we will primarily consider the regime
$N_M\lesssim N_{\footnotesize\textrm{gap}}$, which is more readily
attainable for current experiments.

A remarkable consequence of the system periodicity is that a
lattice of many atoms can form a nearly perfect mirror around
resonance with $1-R_{N_M}{\approx}2\Gamma'/(N_M\GammaD)$, even if
a single atom is mostly absorptive~($\Gamma'>\GammaD$). As shown
below, this effect arises from the excitation of a collective
super-radiant atomic mode whose coupling with the waveguide is strongly
enhanced. This expression reproduces the known result~\cite{fan05,chang07b} that a single emitter~($N_M=1$) can have strong reflectance when $\GammaD/\Gamma'\gg 1$, which can physically occur when atoms are coupled to extremely narrow metallic nanowires~\cite{chang06}. Our result is appealing as it demonstrates that using extremely small guiding structures can be avoided simply by having optical depth as a resource. The collective interaction in our system produces a number of other
interesting phenomena as well. First, the envelope of a
propagating field attenuates through the lattice in a
non-exponential manner, as plotted in figure~\ref{fig:cavity}c across
sites $1<j<N_M$. However, each atom sees the \textit{same},
site-independent local field intensity, given on resonance by
$|E(z_j)/E_0|^2=\frac{\Gamma'^{2}}{(\Gamma'+N_M\GammaD)^2}$, where
$E_0$ is the peak amplitude. The fact that each atom sits
progressively closer to the nodes with increasing $N_M$
suppresses free-space scattering and builds up the large
reflection amplitude. Although the lattice is highly reflective on
resonance, it is also ``dark,'' in that the remaining light is
mostly scattered by the atoms into free space as opposed to
transmitted, $\mathcal{L}_{N_M}\equiv
1-R_{N_M}-T_{N_M}{\gg}T_{N_m}$. The mirror can be made mostly
dispersive~($\mathcal{L}_{N_M}{\ll}T_{N_M}\ll R_{N_M}$) by
operating in a range of detunings where
$N_M\GammaD\gg|\Delta_A|\gg\sqrt{N_M\GammaD\Gamma'}$, at the
expense of needing more atoms to reach a given reflectance.

These collective modes can be leveraged to produce strong coupling
between the fiber and a \textit{single}, specially chosen atom
from within the ensemble. This approach is illustrated in the
``cavity QED'' configuration of figure~\ref{fig:chain}c. As the
nomenclature suggests, here two atomic Bragg mirrors~(at positions
$-N_M\leq j \leq -1$ and $1 \leq j \leq N_M$, for $N_A=2N_M$ total
mirror atoms) form an effective cavity for an impurity atom
located between them at $j=0$. The impurity atom is located a
distance $d_I$ from its nearest neighbors. We will focus on the
geometry where $d_I=3\lambda_A/4$ and $d_M=\lambda_A/2$, such that
the impurity sits at a cavity anti-node and the coupling is
maximized. In analogy to conventional cavity QED, the coupling
between the impurity atom and fiber should be enhanced by the
number of round trips~$\sim N_{A}\GammaD/\Gamma'$ a photon makes
within the cavity.

The spectral properties of this system are illustrated in
figure~\ref{fig:spectra}. Here, we calculate the fields generated by
an impurity atom that is driven from free space, as in
figure~\ref{fig:chain}c. The driving field $\mathcal{E}$ is taken to
be sufficiently weak that atomic saturation can be ignored, with
the atom generating the intra-cavity field profile seen in
figure~\ref{fig:cavity}c. In figures~\ref{fig:spectra}a,b, we plot the
intra-cavity field intensity $I_c$ at the impurity atom position
and the intensity $T_c$ transmitted by either atomic mirror, as a
function of the drive detuning $\Delta_A$. The observed normal
mode splittings suggest that we reach the ``strong coupling''
regime, where the coherent interaction strength between the
impurity atom and cavity mode exceeds their individual dissipative
rates~\cite{sanchez-mondragon83,thompson92,brune96,boca04,miller05,walther06,haroche06}.
As shown in figure~\ref{fig:spectra}c, the splitting between the two
peaks $\Omega_{\pm 1}$ increases as $\Omega_{+1}-\Omega_{-1}\equiv
2g\approx\sqrt{N_A}\GammaD$ for
$N_A{\lesssim}N_{\footnotesize\textrm{gap}}$ and approaches a constant value in the band gap regime $N_A{\gtrsim}N_{\footnotesize\textrm{gap}}$. The normal mode splitting is also illustrated in
figure~\ref{fig:spectra}d, where we allow the resonance frequency of
the impurity atom $\omega_I$ to be separately tuned from that of
the mirror atoms, $\omega_A$.

\subsection{From quantum spin to Jaynes-Cummings model}

While these results can be derived within the transfer matrix
formalism, we now present a more powerful interacting spin model
that elucidates the origin of the strong coupling. The general
field solution of equation~(\ref{eq:fieldsoln}) can be substituted into
the atomic evolution equations, resulting in expressions where the
evolution of atomic coherence $j$, $\dot{\sigma}_{ge}^{j}(t)$,
depends on the coherence of other atoms $k$ at retarded times,
$\sigma_{ge}^{k}(t-|z_j-z_k|/v)$. Further simplification results
if the atomic coherences are slowly varying,
$\sigma_{ge}^j(t-\epsilon)\approx\sigma_{ge}^j(t)e^{i\omega_A\epsilon}$,
and if the characteristic bandwidth $\Delta\omega$ of the dynamics
satisfies $\Delta\omega L/v{\ll}1$, where $L$ is the system
length. In this limit, the photon-mediated dipole-dipole
interactions between atoms are described by a master equation
$\dot{\rho}=-i[H_{dd},\rho]+\mathcal{L}_{dd}[\rho]$ for the atomic
density matrix $\rho$, where
\be H_{dd}=(\GammaD/2)\sum_{j,k}\sin
k_{A}|z_j-z_k|\sigma_{eg}^j\sigma_{ge}^k~\label{eq:Hdd} \ee
and
\be \mathcal{L}_{dd}[\rho]=-(\GammaD/2)\sum_{j,k}\cos
k_{A}|z_j-z_k|\left(\sigma_{eg}^j\sigma_{ge}^k\rho+\rho\sigma_{eg}^j\sigma_{ge}^k-2\sigma_{ge}^k\rho\sigma_{eg}^j\right).\label{eq:Ldd}
\ee
Here $k_A=2\pi/\lambda_A$ is the resonant wavevector, and the sum
on $j,k$ runs over all atoms. The Hamiltonian
characterizes field-mediated dipole exchange between atoms, while
the incoherent evolution $\mathcal{L}_{dd}$ characterizes
cooperative emission. Interestingly, the interactions are infinite
in range and sinusoidal. These features can be qualitatively
understood by noting that a photon emitted by one atom into the
fiber propagates without attenuation until it scatters off a
second atom, and the interaction should be sensitive only to the
relative phases between them. Similar equations have been
previously derived within the more restrictive Born-Markov
approximation~\cite{kien05,kien08,dzsotjan10}. Although the fields
have apparently been eliminated, we note that they can be
reconstructed using equation~(\ref{eq:fieldsoln}). We also include the effects of independent atomic emission into free space at a rate $\Gamma'$ through a separate contribution
$\mathcal{L}_{\scriptsize\textrm{ind}}[\rho]$ to the density
matrix evolution.

Applying the spin model to the cavity QED configuration yields an
interaction Hamiltonian
$H_{dd}=g(\hat{s}^{-}\hat{S}^{+}_{\footnotesize\textrm{cav}}+h.c.)$,
where $g\equiv\GammaD\sqrt{N_A}/2$. Here, we have defined lowering
operators $\hat{s}^{-}=\sigma_{ge}^{(j=0)}$ for the impurity atom
and
$\hat{S}^{-}_{\footnotesize\textrm{cav}}=N_A^{-1/2}\sum_{j>0}(\sigma_{ge}^{j}+\sigma_{ge}^{-j})(-1)^{j}$
for a cavity ``photon'' consisting of a collective spin wave of
the mirror atoms. $\hat{S}^{-}_{\footnotesize\textrm{cav}}$ is a
canonical lowering operator from which other angular momentum
operators can be constructed. These operators together satisfy the
usual angular momentum commutation relations, which can be used to
determine the spectrum of $H_{dd}$.

\begin{figure}[p]
\begin{center}
\includegraphics[width=9cm]{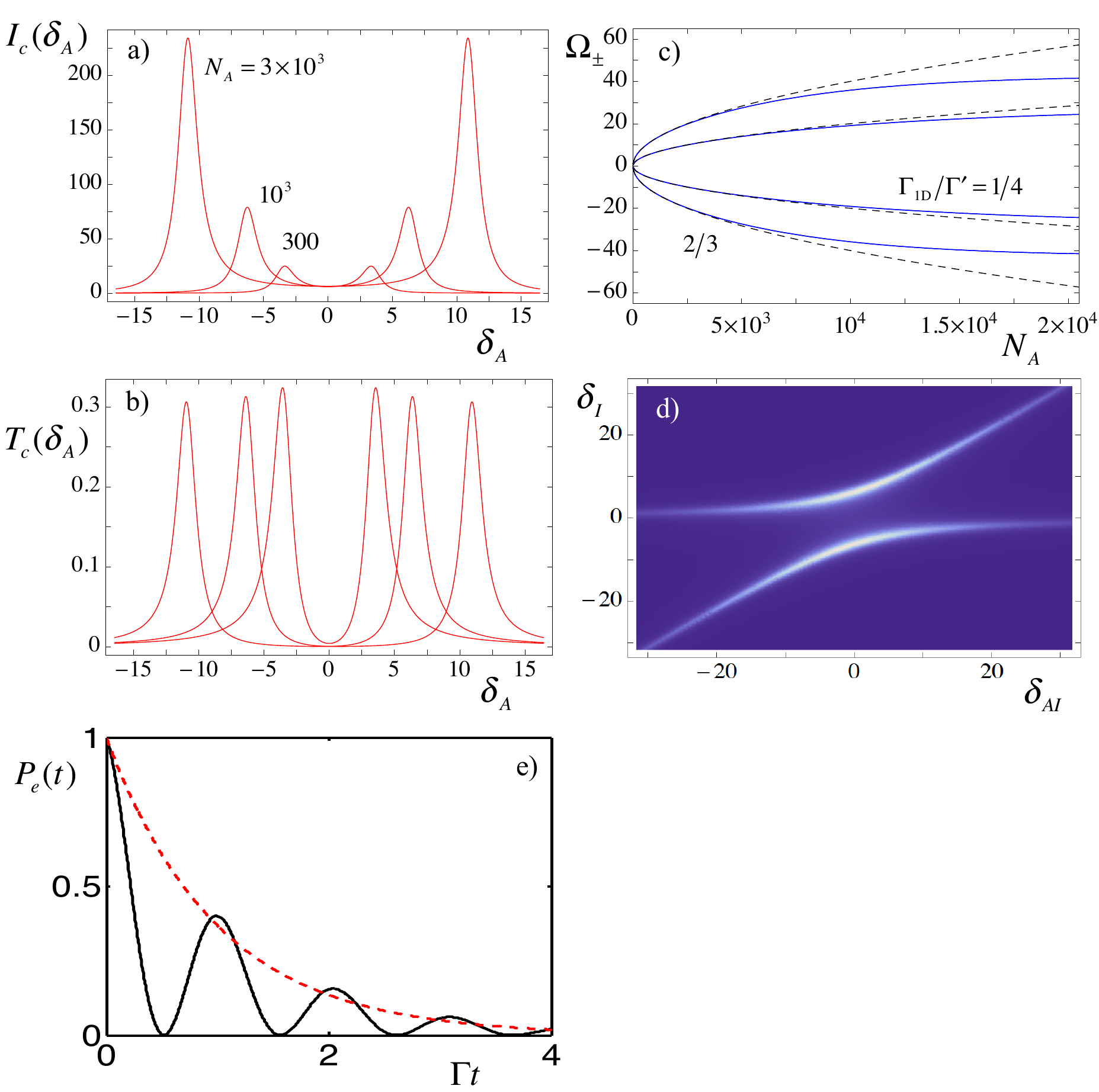}
\end{center}
\caption{\textbf{Strong coupling regime of cavity QED.} Figures a)-d) depict spectra for the cavity configuration of
figure~\ref{fig:chain}c, with $N_A$ total mirror atoms. \textbf{a)} The
impurity atom is driven by an external field $\mathcal{E}$, with
dimensionless detuning $\delta_A=\Delta_A/(\Gamma/2)$ relative to
all of the atoms. The intra-cavity intensity $I_c \equiv
|E_{R}(z=0)|^2$ exhibits a normal mode splitting with peaks at
$\Omega_{\pm}\approx\GammaD\sqrt{N_A}/2$. Here we have chosen
$\GammaD=\Gamma'/4$. \textbf{b)} The intensity $T_c \equiv
|E_{R}(z=z_{N_M})|^2$ transmitted by a single mirror for the same
conditions as in a). $I_c$ and $T_c$ are normalized to the
intensity emitted by a single atom driven by the same amplitude
$\mathcal{E}$ on resonance, absent the atomic mirrors. \textbf{c)} Solid
lines: positions of the normal mode peaks $\Omega_{\pm 1}$ for
$I_c(\delta_A)$ versus atom number, for $\GammaD=\Gamma'/4$ and
$\GammaD=2\Gamma'/3$. The normal mode splitting is
well-approximated by $\Omega_{\pm
1}=\pm\GammaD\sqrt{N_A}/2$~(dashed lines) for atom numbers
$N_A\lesssim N_{\footnotesize\textrm{gap}}$ and saturates for
larger atom number. \textbf{d)} Spectra for the intra-cavity intensity
$I_c$ when the detunings of the mirror atoms and impurity atom are
separately tuned. Here $\delta_I=(\omega_P-\omega_I)/(\Gamma/2)$
denotes the detuning of the impurity atom relative to the probe
beam, while $\delta_{AI}=(\omega_A-\omega_I)/(\Gamma/2)$ denotes
the difference between the mirror and impurity atom resonance
frequencies. \textbf{e)} The population $P_e(t)$ of an initially excited,
single impurity atom inside an atomic cavity~(solid curve), which
exhibits vacuum Rabi oscillations as the excitation is reversibly
exchanged with a spin wave comprising the mirror atoms at a rate
$g=\GammaD\sqrt{N_A}/2$. We have used $\GammaD=\Gamma'/4$ and $N_A=900$ atoms. For
comparison, the dashed red curve shows the spontaneous emission
decay of a single excited atom absent the cavity,
$P_e(t)=e^{-\Gamma t}$.\label{fig:spectra}}
\end{figure}

In particular, the dipole-dipole interaction splits the nominal
degeneracy between the state where $n_{\footnotesize\textrm{cav}}$
excitations are contained in the cavity spin mode,
$\ket{g,n_{\footnotesize\textrm{cav}}}\propto(\hat{S}^{+}_{\footnotesize\textrm{cav}})^{n_{\footnotesize\textrm{cav}}}\ket{g}^{\otimes
(N_A+1)}$, and the state with $n_{\footnotesize\textrm{cav}}-1$ excitations in the spin mode
and one excitation in the impurity atom,
$\ket{e,n_{\footnotesize\textrm{cav}}-1}\propto\hat{s}^{+}(\hat{S}^{+}_{\footnotesize\textrm{cav}})^{n_{\footnotesize\textrm{cav}}-1}\ket{g}^{\otimes
(N_A+1)}$. The new eigenstates are given by
$\ket{\pm,n_{\footnotesize\textrm{cav}}}=(\ket{g,n_{\footnotesize\textrm{cav}}}\pm\ket{e,n_{\footnotesize\textrm{cav}}-1})/\sqrt{2}$,
with corresponding energies
$\Omega_{{\pm},n_{\footnotesize\textrm{cav}}}{\approx}\pm
g\sqrt{n_{\footnotesize\textrm{cav}}}$ in the regime of small
excitation number $n_{\footnotesize\textrm{cav}}\ll N_A$~(where
saturation is negligible and the mirror atom excitations are
nearly bosonic). This excitation spectrum is intrinsically
anharmonic and identical to that of the Jaynes-Cummings model
describing a single two-level atom coupled to a conventional
cavity~\cite{jaynes63}. The linear case of
$n_{\footnotesize\textrm{cav}}=1$ yields $\Omega_{{\pm},1}=\pm
\GammaD\sqrt{N_A}/2$, reproducing the splitting in the spectrum
observed in figure~\ref{fig:spectra} for $N_M\lesssim
N_{\footnotesize\textrm{gap}}$. This mapping onto the Jaynes-Cummings model is important in two respects. First, its nonlinearity is known to be critical to various tasks in quantum information processing based on cavity QED~\cite{kimble08}. Second, the ability to reduce our
\textit{a priori} multi-mode atomic ensemble to a single mode enables relatively simple dynamics and exact solutions, which are generally absent in the multi-mode case~\cite{hafezi11}. This feature enables tasks in quantum information to be executed with reduced errors and high fidelity.

The dissipation rates of the cavity configuration can be similarly
characterized, by writing
$\mathcal{L}_{dd}[\rho]=-(\GammaD/2)(\hat{s}^{+}\hat{s}^{-}\rho+\rho
\hat{s}^{+}\hat{s}^{-}-2\hat{s}^{-}\rho\hat{s}^{+})-(N_A\GammaD/2)(\hat{S}^{+}_{\footnotesize\textrm{rad}}\hat{S}^{-}_{\footnotesize\textrm{rad}}\rho+\rho
\hat{S}^{+}_{\footnotesize\textrm{rad}}\hat{S}^{-}_{\footnotesize\textrm{rad}}-2\hat{S}^{-}_{\footnotesize\textrm{rad}}\rho\hat{S}^{+}_{\footnotesize\textrm{rad}})$.
Here
$\hat{S}^{-}_{\footnotesize\textrm{rad}}=N_A^{-1/2}\sum_{j>0}(\sigma_{ge}^{j}-\sigma_{ge}^{-j})(-1)^{j+1}$
is a lowering operator for a spin wave of the mirror atoms with
super-radiant emission. While angular momentum operators
constructed from $\hat{S}^{-}_{\footnotesize\textrm{rad}}$ obey
canonical commutation relations amongst themselves, the two spin
waves associated with $\hat{S}^{-}_{\footnotesize\textrm{cav}}$
and $\hat{S}^{-}_{\footnotesize\textrm{rad}}$ have non-trivial
commutation relations between them. For example,
$\hat{S}^{-}_{\footnotesize\textrm{rad}}\ket{1_{\footnotesize\textrm{cav}}}=0$, indicating
that a single cavity excitation does not emit into the waveguide.
Thus, its decay rate is given by the single-atom emission rate
into free space, $\kappa=\Gamma'$. The origin of the sub-radiance
is destructive interference between the light emitted by pairs of
mirror atoms on each side of the impurity~(say $\pm j$), as
illustrated in figure~\ref{fig:chain}c. Here, one sees that each
atom in the pair $\pm j$ has the same phase $(-1)^{j}$. However,
because they are spaced an odd multiple of $\lambda_A/2$ apart,
their radiation into the waveguide perfectly cancel. This effect
also gives rise to the high transmitted intensity $T_c$ of light
produced by the impurity atom~(figure~\ref{fig:spectra}b). Interestingly, applying
$\mathcal{L}_{dd}$ to the spin wave of only a single mirror~(say
$1\leq j \leq N_M$) reveals that such a state is maximally
super-radiant~\cite{kien08}, giving rise to its high reflectance.
Likewise, one can show that the decay rate of the state
$\ket{e,0_{\footnotesize\textrm{cav}}}$~(an excited impurity atom) is
$\Gamma=\GammaD+\Gamma'$.

In analogy with cavity QED, one can associate various figures of
merit to $g,\kappa,\Gamma$. For example, the enhanced coupling to
the cavity mode by the impurity atom is characterized by the
cooperativity $C\equiv
\frac{g^2}{\kappa\Gamma}=\frac{\GammaD}{\GammaD+\Gamma'}\frac{N_A\GammaD}{\Gamma'}$.
Note that $\frac{\GammaD}{\GammaD+\Gamma'}$ represents the
single-atom coupling efficiency to the waveguide, while
$\frac{N_A\GammaD}{\Gamma'}$ is proportional to the cavity
finesse~(figure~\ref{fig:cavity}b). Surprisingly, our
results also show that with modest atom numbers one can reach the
strong coupling regime $g>\kappa,\Gamma$, where an impurity atom
can emit and then re-absorb the same photon~(the so-called vacuum
Rabi oscillations~\cite{jaynes63,sanchez-mondragon83}).

In contrast to the transfer matrix formalism, our interacting spin
model is ideal to studying the system dynamics in the quantum
regime. In figure~\ref{fig:spectra}e, we plot the analytic solution for
the time evolution $\dot{\rho}$ starting with an initially excited
impurity atom~($\rho=\ket{e,0_{\footnotesize\textrm{cav}}}\bra{e,0_{\footnotesize\textrm{cav}}}$ at $t=0$). Rabi
oscillations of the impurity excited state population are clearly
visible in the case of $N_A=900$ atoms and
$\GammaD=\Gamma'/4$~($g=3\Gamma,\kappa=0.8\Gamma$). This feature
can be viewed in the dressed-state picture as an interference
effect between the states $\ket{\pm,1_{\footnotesize\textrm{cav}}}$ which make up the initial
state.

The strong coupling regime for a single impurity atom can be reached with very low finesse for the atomic mirrors~(\textit{e.g.}, $F\sim 590$ in figure~\ref{fig:spectra}a with $N_A=3000$ atoms, while $F\sim 175 $ in figure~\ref{fig:spectra}e with $N_A=900$ atoms; see figure~\ref{fig:cavity}b). By contrast, for a conventional Fabry-Perot cavity with dielectric mirrors, strong coupling requires finesse $F\gtrsim 10^5$~\cite{miller05}. In fact, the decay rate $\kappa$ relevant to strong coupling with atomic mirrors as in figure~\ref{fig:chain}c is that of the sub-radiant mode of the atomic chain~($\kappa=\Gamma'$). The highly dispersive nature of these atoms causes $\kappa$ to be much smaller than the conventional cavity decay rate $\kappa_c=v\pi/FL_{\footnotesize\textrm{eff}}$, where $L_{\footnotesize\textrm{eff}}$ is the effective cavity length and $F$ is the finesse set by the mirror reflectivity. In this regard, note that the sub-radiant mode is not relevant to the dielectric coating of a conventional high-reflectivity mirror because of the rapid relaxation of the polarizability of the dielectric elements. Furthermore, although we have focused on the case of perfect filling of the atomic mirror lattice sites, it is clear from the infinite-range,
sinusoidal form of the interactions that these effects are quite robust to filling imperfections and rely solely on the system periodicity.

\subsection{Building blocks for scalable quantum information processing}

\begin{figure}[p]
\begin{center}
\includegraphics[width=14cm]{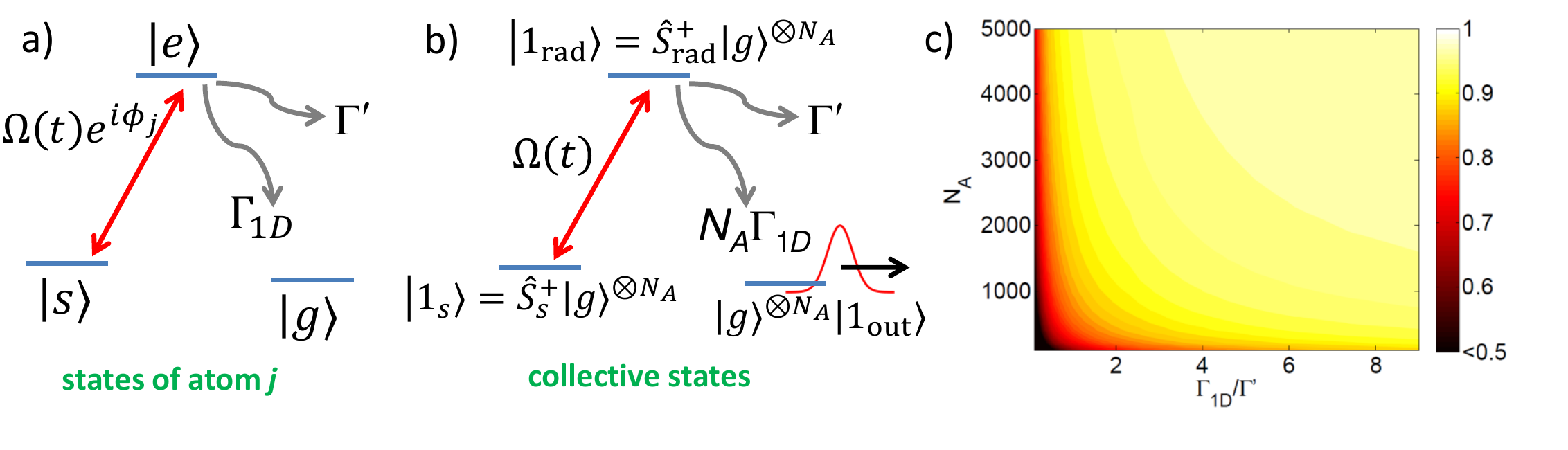}
\end{center}
\caption{\textbf{Building blocks for quantum information processing.} \textbf{a)} Three-level structure of a single atom. The ground state $\ket{g}$ is coupled via waveguide modes to excited state $\ket{e}$, while a meta-stable state $\ket{s}$ is de-coupled from the waveguide but can be coupled to $\ket{e}$ through an external control field $\Omega(t)e^{i\phi_j}$. The excited state decays into free space and the waveguide at rates $\Gamma',\GammaD$, respectively. \textbf{b)} The collective states of the cavity mirror atoms used to efficiently map between atomic excitations and propagating photons. With a proper choice of driving phases $\phi_j$, the external field $\Omega(t)$ couples a meta-stable spin excitation
$\ket{1_s}$ in the mirror atoms to a super-radiant, excited-state spin wave $\ket{1_{\footnotesize\textrm{rad}}}$. This state emits into the waveguide at an enhanced rate $N_{A}\GammaD$, generating an outgoing photon
$\ket{1_{\footnotesize\textrm{out}}}$ with high probability. The time-reversed process enables an incoming photon to be converted to a meta-stable spin excitation. \textbf{c)} Fidelity for quantum state transfer between two impurities $p,q$ in a cavity formed by $N_A$ mirror atoms exterior
to the impurities~(see figure~\ref{fig:chain}d). The fidelity of the adiabatic transfer process is numerically optimized as functions of the single-atom coupling strength to the waveguide~($\GammaD/\Gamma'$) and mirror atom number $N_A$.\label{fig:quantum}}
\end{figure}

Here, we describe how our cavity QED system can be used to realize the basic building blocks for scalable quantum information processing. As with other atom-light interfaces~\cite{duan08,kimble08}, the
utility of the present system is greatly extended by
introducing an atomic meta-stable state $\ket{s}$~(see figure~\ref{fig:quantum}a), which is
decoupled from the fiber modes due to an orthogonal dipole
orientation, but which can be coupled to $\ket{e}$ through a
time-dependent external optical field with strength
$\Omega(t)e^{i\phi_j}$ for atom $j$.
Here we assume that the Rabi amplitude $\Omega(t)$ is identical
for all atoms, but we allow for the possibility of a varying phase
$\phi_j$, which can be used to couple to selective spin waves. As
we now describe, this coupling can be used to faithfully map the
quantum states of propagating waveguide photons into meta-stable spin excitations and
back to provide a long-lived quantum memory. The coupling also enables these meta-stable spin excitations to be loaded into the cavity, which allows for quantum logic and other non-classical operations to be performed.

We first investigate the mapping of a single, meta-stable spin wave excitation in the atom mirrors to an outgoing photon. The spin wave of interest is characterized by
the operator
$\hat{S}^{-}_{\footnotesize\textrm{s}}=N_A^{-1/2}\sum_{j>0}(\sigma_{gs}^{j}-\sigma_{gs}^{-j})(-1)^{j+1}$,
such that the initial state of the mirror atoms is given by
$\ket{1_s}\equiv\hat{S}^{+}_{\footnotesize\textrm{s}}\ket{g}^{\otimes
N_A}$. The impurity atom is assumed to be in state $\ket{s}$ and undriven by external fields, so that it does not participate in this process. The external field $\Omega(t)e^{i\phi_j}$ driving the mirror atoms couples
$\ket{1_s}$ to the super-radiant, excited-state spin wave
$\ket{1_{\footnotesize\textrm{rad}}}\equiv\hat{S}^{+}_{\footnotesize\textrm{rad}}\ket{g}^{\otimes
N_A}$ if the driving phase for the atoms is equal, say
$\phi_j=0$, as shown in figure~\ref{fig:quantum}b. Note that
$\ket{1_{\footnotesize\textrm{rad}}}$ couples with maximum
spontaneous emission rate $N_A\GammaD$ into the waveguide,
compared to $\Gamma'$ into free space. This feature of
$\ket{1_{\footnotesize\textrm{rad}}}$ enables efficient mapping of
the meta-stable spin wave $\ket{1_s}$ into an outgoing photon
$\ket{1_{\footnotesize\textrm{out}}}$ in the waveguide. Generally, a proper choice of $\Omega(t)$ can
produce an outgoing photon of any desired shape within a bandwidth
${\lesssim}N_A\GammaD$, and with an error probability of
$\Gamma'/N_A\GammaD$ due to free-space leakage~\cite{gorshkov06,gorshkov07,chang07b}. It should be noted that this outgoing photon is split equally into left- and right-propagating modes, due to the symmetry of the super-radiant spin wave.

By time reversal symmetry~\cite{cirac97,gorshkov06,gorshkov07}, it also follows that an incoming photon in the waveguide~(in an equal superposition of left- and right-propagating modes) of bandwidth
${\lesssim}N_A\GammaD$ can be mapped into a spin excitation $\ket{1_s}$ starting from an initial atomic mirror state $\ket{g}^{\otimes N_A}$  with the same error $\Gamma'/N_A\GammaD$. These mappings to and from $\ket{1_s}$ thus provide an efficient interface between propagating fields and the atomic ensemble.

In addition, $\ket{1_s}$ can be efficiently coupled to a cavity
excitation
$\ket{1_{\footnotesize\textrm{cav}}}\equiv\hat{S}^{+}_{\footnotesize\textrm{cav}}\ket{g}^{\otimes
N_A}$ by choosing a different relative phase for the control
field, \textit{e.g.}, $\phi_j=0$ for $j>0$ and $\phi_j=\pi$ for
$j<0$. These separate processes of mapping $\ket{1_s}$ between
outgoing photons and cavity excitations is necessary in our system
because the cavity excitation is nominally de-coupled from the
waveguide~(being maximally sub-radiant). From here, however, our
system behaves identically to a conventional cavity QED system
governed by the Jaynes-Cummings model. In particular, one can apply to our system existing
information processing protocols such as for conditional quantum logic between two photons~\cite{duan04} or impurity atoms~\cite{pellizzari95}, or quantum state transfer between two such atoms~\cite{pellizzari95}.

As a specific example, we analyze how our system can serve as an efficient quantum information bus
between two distant impurity atoms within the same chain, in analogy to the case of two atoms in a conventional cavity~\cite{pellizzari95}. One
possible configuration is illustrated in figure~\ref{fig:chain}d,
where two well-separated impurity atoms $p,q$ are initially
embedded in a long chain of mirror atoms in state $\ket{g}$. To
facilitate information transfer, the mirror atoms between $p,q$
are first flipped into the transparent meta-stable state $\ket{s}$, and thus do not participate in the process. Through this operation, the impurity atoms are loaded
into a new, common cavity mode, which is defined by the mirror
atoms external to $p,q$ and which mediates coherent information transfer
between the two impurities. The objective of the state transfer process is to map an arbitrary quantum bit encoded in the states $s,g$ from $p$ to $q$, \textit{i.e.}, $(
c_1\ket{s_p}+c_2\ket{g_p})\ket{g_q}\rightarrow \ket{g_p}(
c_1\ket{s_p}+c_2\ket{g_p})$. We assume that the impurity atoms can be driven by individual external control fields $\Omega_{p,q}(t)$ on the $\ket{s}$-$\ket{e}$ transition. These control beams clearly have no effect on the state $\ket{g_p,g_q}$, and we describe how a proper choice of the control fields yields $\ket{s_p,g_q}\rightarrow\ket{g_p,s_q}$ to enable the desired transfer of an arbitrary superposition. As noted in Ref.~\cite{pellizzari95}, there exists an instantaneous dark eigenstate of the system Hamiltonian given by $\ket{D(t)}\propto g\Omega_q(t)\ket{s_p,g_q,0_{\footnotesize\textrm{cav}}}+g\Omega_p(t)\ket{g_p,s_q,0_{\footnotesize\textrm{cav}}}-\Omega_p(t)\Omega_q(t)\ket{g_p,g_q,1_{\footnotesize\textrm{cav}}}$. Note that the state $\ket{s_p,g_q}$~($\ket{g_p,s_q}$) corresponds to $\ket{D(t)}$ in the limit where $\Omega_p=0$~($\Omega_q=0$). The desired transformation can thus be achieved through adiabatic passage using a pulse sequence that leads from $\Omega_{p}(t=0)=0$ to $\Omega_{q}(T)=0$ over a time $T\gg 1/g,1/\Omega_0$, where $\Omega_0$ is the characteristic amplitude of $\Omega_{p,q}$. Since $\ket{D(t)}$ and $\ket{g_p,g_q}$ have the same energy, coherence of an arbitrary superposition is maintained throughout the process.

In figure~\ref{fig:quantum}c, we plot the fidelity of the transformation $\ket{s_p,g_q}\rightarrow\ket{g_p,s_q}$ as functions of $\Gamma_{1D}/\Gamma'$ and mirror atom number $N_A$~(determined by the number of atoms external to $p,q$). Note that this represents the lower bound on the transfer fidelity of an arbitrary state, as the state $\ket{g_p,g_q}$ is unaffected by the pulse sequence. Here we have chosen the pulse sequence $\Omega_{p}(t)=\Omega_{0}\sin\;\frac{\pi t}{2T}$ and $\Omega_{q}(t)=\Omega_{0}\cos\;\frac{\pi t}{2T}$~($0\leq t \leq T$) with overall pulse duration $T=50/g$, and we have optimized $\Omega_0$ by numerically solving our spin model. The optimized error of the state transfer process depends on the cavity cooperativity factor approximately as $\sim 1/\sqrt{C}$, which reflects an optimized balance between dissipation of the cavity excitation component of the dark state $\ket{D}$ and non-adiabatic transitions out of the dark state. A unique feature of our system compared to a conventional cavity is that the coupling strength $g$ does not decrease with increasing cavity mode volume~(\textit{i.e.}, increasing separation between $p,q$), so that the amount of time required for the adiabatic process remains constant.

\section{Discussion}

We have described a novel technique to realize and manipulate
strong photon-atom coupling using cold atoms trapped near a
tapered nanofiber~\cite{nayak07,vetsch10,lacroute11}. Our approach
combines concepts from cavity QED, collective enhancement in
atomic ensembles, and tight focusing of optical fields to achieve
the strong coupling regime using relatively modest resources, and
can be used for scalable quantum information processing.

Thus far, we have investigated the case of a single excitation,
but we anticipate that nonlinear and many-body behavior involving
atoms and photons~\cite{chang08b,gorshkov10,kiffner10,shahmoon11}
will be an interesting topic for further exploration. For example,
this system may allow for the experimental study of quantum spin
models with infinite-range interactions~\cite{lipkin65}. This
system could also stimulate interesting studies into the role of
atomic disorder in field propagation~\cite{kirkman84} and its
interplay with interactions~\cite{giamarchi88}. Furthermore, the ability to map cavity excitations onto long-lived atomic quantum memories and subsequently to output fields can enable the generation of non-classical, many-photon states~(\textit{e.g.}, using the techniques of Ref.~\cite{law96}), which find applications in areas such as enhanced quantum metrology and sensing~\cite{dowling08}.

Finally, although
we have focused on a simple fiber geometry here, we envision that
an even richer set of phenomena can occur when the waveguide
itself is allowed to have structure, such as in a photonic crystal
nanowire~\cite{eichenfield09}. Here, for example, one could
engineer the dispersion relations~\cite{chan09} to provide
commensurate wavevectors between the trapping and resonant light
and large single-atom coupling efficiencies
$\GammaD/\Gamma$. It should also be feasible to tailor the
structure to introduce selective phase slips, which could define
impurity atom sites and create more exotic interactions with
broken translational invariance. Moreover, these structures could
contain additional degrees of freedom, such as mechanical
modes~\cite{eichenfield09,chan11}, to which atoms can
provide a quantum interface~\cite{hammerer09}.

\begin{acknowledgements}
The authors thank O. Painter, A. Goban, D. Ding, M.
Pototschnig, and J.I. Cirac for valuable discussions. DEC acknowledges support
from Fundaci\'{o} Privada Cellex Barcelona. LJ acknowledges
support from the Sherman Fairchild Foundation and the NBRPC (973
program) 2011CBA00300 (2011CBA00301). AVG acknowledges support
from the Lee A. DuBridge Foundation. Funding at Caltech is
provided by the Institute for Quantum Information and Matter, an
NSF Physics Frontiers Center with support of the Gordon and Betty
Moore Foundation, by NSF Grant PHY0652914, by the DoD NSSEFF
program, and by the AFOSR MURI for Quantum Memories.
\end{acknowledgements}

\end{document}